\expandafter\def\csname r@sec-twop\endcsname{2\null}
\documentstyle[12pt]{article}
\title{Some eigenstates for a model associated with solutions
       of tetrahedron equation.\\
       V.~Two cases of string superposition}
\author{I.G.~Korepanov}
\date{May 1997}
\def\be{\begin{equation}}
\def\ee{\end{equation}}
\makeatletter
\long\def\@makecaption#1#2{\vskip 10\p@ \hbox to\hsize{\hfil#1\hfil}}
\makeatother
\begin{document}
\maketitle

\begin{abstract}
In paper IV (solv-int/9704013) we have considered a string
living in the infinite lattice
that was, in a sense, generated by a ``particle''.
Here we show how to construct multi-string eigenstates
generated by several particles. It turns out that, at least
in some cases, this allows us to bypass the difficulties
of constructing multi-particle states. We also present and discuss
the ``dispersion relations'' for our particles--strings.
\end{abstract}

\section*{Introduction}

Let us recall that we have introduced in paper~\cite{I} some
``one-particle'' eigenstates for the model based upon solutions
of the tetrahedron equation. In the same paper, we have also
constructed some ``two-particle'' states. However, some
special condition arised in this construction, and the
superposition of two {\em arbitrary\/} one-particle states
was not achieved. Even the ``creation operators'' of paper~\cite{II}
did not give a clear answer concerning multi-particle states.

On the other hand, in paper~\cite{IV} we have brought in correspondence
to a one-particle state some new state that could be
called ``one open string''. It was done using some special ``kagome
transfer matrix''. Here we will show that the superposition
of such one-string states is easier to construct, because of
degeneracy of kagome transfer matrix: it turns into zero the
``obstacles'' that hampered constructing of multi-particle states.

The scheme of string---particle ``marriage'' in~\cite{IV} was
as follows: take a one-particle state from \cite{I,II}, and apply
to it a kagome transfer matrix with boundary conditions corresponding
to the presence of two string tails in the infinity, e.g.\ like this:
\allowbreak
\unitlength=1mm
\linethickness{0.4pt}
\begin{picture}(25.00,4.50)
\put(25.00,2.00){\oval(50.00,4.00)[l]}
\end{picture}\,.

In this paper, we are going to complicate this scheme in the following
way: the boundary conditions will correspond to the presence
of an even number
of string tails at the infinity, and instead of a one-particle
state, we will use some special multi-particle vector~$\Psi$.
Its peculiarity will be in the fact that $\Psi$ is {\em no longer
an eigenstate\/} of the hedgehog transfer matrix~$T$ defined in~\cite{I}.
Instead, it will obey the condition
\be
T\Psi = \lambda\Psi + \Psi',
\label{V-int-1}
\ee
where $\lambda={\rm const}$, and $\Psi'$ is annulated by the
kagome transfer matrix of paper~\cite{IV} which we will
denote~$K$.

Recall that we have defined $T$ in such a way that its degrees
could be described geometrically as ``oblique slices''
of the cubic lattice.
The transfer matrix~$T$ can be passed through the transfer matrix~$K$:
\be
TK=KT,
\label{V-int-2}
\ee
the boundary conditions (such as the number and form of tails
at the infinity) for $K$ being intact. Define vector~$\Phi$ as
$$
\Phi=K\Psi.
$$
This together with (\ref{V-int-1}) and (\ref{V-int-2}) gives
\be
T\Phi=\lambda\Phi,
\label{V-int-4}
\ee
exactly as needed for an eigenvector.

We also present in this paper
the ``dispersion relations'' for our particles--strings
in a workable form---something that was missing in
papers~\cite{I,II}.

\section{Eigenvectors of the ``several open strings'' type for the infinite
lattice}
\label{secV-1}

Let there be $n$ one-particle amplitudes $\varphi_{\ldots}^{(1)}, \ldots,
\varphi_{\ldots}^{(n)}$ of the same type as those described
in the work~\cite{I}. Let us compose an ``$n$-particle vector''
$\Psi$, i.e.\ put in correspondence to each unordered
$n$-tuple of vertices $A^{(1)},\ldots,A^{(n)}$ of the kagome lattice
the symmetrized amplitude in the following way:
\be
\psi_{A^{(1)},\ldots,A^{(n)}} = \sum_s \varphi_{A^{s(1)}}^{(1)}
\ldots \varphi_{A^{s(n)}}^{(n)},
\label{V-1-1}
\ee
where $s$ runs through the group of all permutations
of the set $\{1,\ldots,n\}$.

As for the boundary conditions for the transfer matrix~$K$ described
in the Introduction, let us assume that there are exactly $2n$ string tails,
and they all go in positive directions,
that is between the east and the north.
Thus, in each of the points $A^{(1)},\ldots,A^{(n)}$ a string is
created, and they are not annihilated.

The vector (\ref{V-1-1}) is not an eigenvector of transfer matrix~$T$
due to problems arising when two or more points $A^{(k)}$ get close
to one another. Nevertheless, the vector $\Phi=K\Psi$ {\em is\/}
an eigenvector, because for it those problems disappear due to the
simple fact:
{\em creation of two or more strings within one triangle of the kagome
lattice is geometrically forbidden}.

\section{Eigenvectors of the ``closed string'' type for the infinite
lattice}
\label{secV-2}

In this section, we will put in correspondence
to each unordered pair of vertices
of the infinite kagome lattice an ``amplitude'' $\Psi_{AB}$
according to the following rules. If one of the vertices,
say $A$, {\em precedes\/} the other one, say $B$, in the sense
that they can be linked by a path---a broken line---going along
lattice edges in positive directions, namely northward, eastward,
or to the north-east, then let us put
\be
\Psi_{AB}= \varphi_A \psi_B - \psi_A \varphi_B,
\label{V-2-1}
\ee
where $\varphi_{\ldots}$ and $\psi_{\ldots}$ are two one-particle
amplitudes of the same type as in paper~\cite{I}.
In the case if vertices $A$ and $B$ cannot be joined by a path
of such kind, let us put
$$
\Psi_{AB}=0.
$$

The values $\Psi_{AB}$ are components of the vector~$\Psi$
that belong to the two-particle subspace of tensor product of
two-dimensional spaces situated in all kagome lattice vertices.
What prevents $\Psi_{AB}$ from being an eigenvector of the hedgehog
transfer matrix is discrepancies arising near those pairs $A,B$
that lie at the ``border'' between such pairs where one of the
vertices precedes the other (so to speak, ``the interval $AB$
is timelike''),
and such pairs where it does not (``the interval $AB$
is spacelike'').

Those discrepancies, however, disappear for the vector $\Phi=K\Psi$,
where $K$ is the kagome transfer matrix described in the Introduction
with the boundary conditions reading {\em no string tails at the
infinity}. This is because if a string cannot, geometrically,
be created at the point $A$ (or $B$) and then annihilated
at the point $B$ (or $A$), then the amplitude $\Psi_{AB}$
doesn't influence at all the vector~$\Phi$.
The only thing that remains to be checked for (\ref{V-int-4})
to hold is a situation where $A$ and $B$ are in the same kagome lattice
triangle that will be turned inside out by one of the hedgehogs
of transfer matrix~$T$, as in Figure~\ref{figV-1}.
\begin{figure}[ht]
\begin{center}
\unitlength=1mm
\linethickness{0.4pt}
\begin{picture}(65.00,33.00)
\put(3.00,12.00){\line(1,0){20.00}}
\put(23.00,12.00){\line(0,1){20.00}}
\put(23.00,32.00){\line(-1,-1){20.00}}
\put(30.00,17.00){\vector(1,0){6.00}}
\put(43.00,2.00){\line(0,1){20.00}}
\put(43.00,22.00){\line(1,0){20.00}}
\put(63.00,22.00){\line(-1,-1){20.00}}
\put(3.00,12.00){\circle*{1.00}}
\put(23.00,12.00){\circle*{1.00}}
\put(23.00,32.00){\circle*{1.00}}
\put(43.00,22.00){\circle*{1.00}}
\put(63.00,22.00){\circle*{1.00}}
\put(43.00,2.00){\circle*{1.00}}
\put(13.00,22.00){\vector(1,1){1.00}}
\put(23.00,22.00){\vector(0,1){1.00}}
\put(13.00,12.00){\vector(1,0){1.00}}
\put(43.00,12.00){\vector(0,1){1.00}}
\put(53.00,12.00){\vector(1,1){1.00}}
\put(53.00,22.00){\vector(1,0){1.00}}
\put(1.00,12.00){\makebox(0,0)[rc]{$A$}}
\put(25.00,32.00){\makebox(0,0)[lc]{$B$}}
\put(41.00,2.00){\makebox(0,0)[rc]{$B'$}}
\put(65.00,22.00){\makebox(0,0)[lc]{$A'$}}
\end{picture}
\end{center}
\caption{}
\label{figV-1}
\end{figure}
Acting in the same manner as in Section~\ref{sec-twop}
of work~\cite{I}, write
\be
\pmatrix{ \varphi_{A'} \cr \varphi_{B'} }=
\pmatrix{\alpha & \beta \cr \gamma & \delta}
\pmatrix{ \varphi_A \cr \varphi_B }, \qquad
\pmatrix{ \psi_{A'} \cr \psi_{B'} }=
\pmatrix{\alpha & \beta \cr \gamma & \delta}
\pmatrix{ \psi_A \cr \psi_B },
\label{V-2-3}
\ee
where
\be
\alpha=-\delta, \qquad \alpha\delta-\beta\gamma=-1.
\label{V-2-4}
\ee
It follows from the formulas (\ref{V-2-3}) and (\ref{V-2-4}) that
$$
\varphi_A \psi_B-\varphi_B \psi_A=
\varphi_{B'} \psi_{A'}-\varphi_{A'} \psi_{B'},
$$
i.e.\
$$
\Psi_{AB}=\Psi_{B'A'},
$$
exactly what was needed to comply with the fact that an
$S$-operator-hedgehog acts as a unity operator in the
two-particle subspace.

\section{Dispersion relations}
\label{V-sec-disp}

The constructed eigenvectors of transfer matrix~$T$ are of course
eigenvectors for translation operators through periods of kagome
lattice as well. Let us consider here relations between
the corresponding eigenvalues, starting from the simplest
one-particle eigenstate.

Consider once again some triangle $ABC$ of the kagome lattice,
and its image $A'B'C'$ under the action of $S$-matrix-hedgehog,
as in Figure~\ref{figV-3-1}.
\begin{figure}[ht]
\begin{center}
\unitlength=1.00mm
\special{em:linewidth 0.4pt}
\linethickness{0.4pt}
\begin{picture}(65.00,32.50)
\put(3.00,12.00){\line(1,0){20.00}}
\put(23.00,12.00){\line(0,1){20.00}}
\put(23.00,32.00){\line(-1,-1){20.00}}
\put(30.00,17.00){\vector(1,0){6.00}}
\put(43.00,2.00){\line(0,1){20.00}}
\put(43.00,22.00){\line(1,0){20.00}}
\put(63.00,22.00){\line(-1,-1){20.00}}
\put(3.00,12.00){\circle*{1.00}}
\put(23.00,12.00){\circle*{1.00}}
\put(23.00,32.00){\circle*{1.00}}
\put(43.00,22.00){\circle*{1.00}}
\put(63.00,22.00){\circle*{1.00}}
\put(43.00,2.00){\circle*{1.00}}
\put(1.00,12.00){\makebox(0,0)[rc]{$A$}}
\put(25.00,32.00){\makebox(0,0)[lc]{$C$}}
\put(41.00,2.00){\makebox(0,0)[rc]{$C'$}}
\put(65.00,22.00){\makebox(0,0)[lc]{$A'$}}
\put(41.00,22.00){\makebox(0,0)[rc]{$B'$}}
\put(25.00,12.00){\makebox(0,0)[lc]{$B$}}
\end{picture}
\end{center}
\caption{}
\label{figV-3-1}
\end{figure}
Let us write out some relations of the type (\ref{V-2-3}), namely
\be
\pmatrix{\varphi_{A'} \cr \varphi_{B'}}=
\pmatrix{a&b\cr c&d} \pmatrix{\varphi_A \cr \varphi_B},
\label{V-3-1}
\ee
\be
\pmatrix{\varphi_{B'} \cr \varphi_{C'}}=
\pmatrix{\tilde a&\tilde b\cr \tilde c&\tilde d}
\pmatrix{\varphi_B \cr \varphi_C},
\label{V-3-2}
\ee
where $\varphi_{\ldots}$ is any one-particle vector, and the numbers
$a, \ldots ,\tilde d$ satisfy conditions of type~(\ref{V-2-4}), i.e.\
$$
\matrix{ a=-d,& \qquad ad-bc=-1, \cr
\tilde a=-\tilde d,& \qquad \tilde a\tilde d-\tilde b\tilde c=-1.}
$$
 From (\ref{V-3-1}) follows
\be
{\varphi_B\over \varphi_{B'}}=
{-a(\varphi_A/\varphi_{A'})+1\over (\varphi_A/\varphi_{A'})-a},
\label{V-3-3}
\ee
and from (\ref{V-3-2}) follows
$$
{\varphi_C\over \varphi_{C'}}=
{-\tilde a(\varphi_B/\varphi_{B'})+1
\over (\varphi_B/\varphi_{B'})-\tilde a}.
$$
Surely, the numbers $a$ and $\tilde a$ depend on an
$S$-operator-hedgehog. On the other hand, this latter is parameterized
by exactly two parameters. So, it seems that it makes sense to take
$a$ and $\tilde a$ as those parameters.

We can take for eigenvalue of the hedgehog transfer matrix~$T$
either $\varphi_{A'}/\varphi_A$, or $\varphi_{B'}/\varphi_B$,
or $\varphi_{C'}/\varphi_C$.
These variants correspond, strictly speaking, to different definitions
of~$T$, but each of them is consistent with the requirement that
the degrees of~$T$ must be represented graphically as ``oblique
layers'' of cubic lattice (the difference being that, with the three
different definitions, the action of transfer matrix~$T$ corresponds
to the shifts through cubic lattice periods along three different axes).
Our goal is to express the eigenvalues of translation operators
acting within the kagome lattice for a given one-particle state
through, say, $\varphi_{A'}/\varphi_A$.

If we speak about translation through one lattice
period {\em to the right\/}
in the sense of Figures \ref{figV-3-1} and~\ref{figV-3-2},
\begin{figure}[ht]
\begin{center}
\unitlength=1mm
\linethickness{0.4pt}
\begin{picture}(50.00,50.00)
\put(25.00,0.00){\line(0,1){50.00}}
\put(0.00,25.00){\line(1,0){50.00}}
\put(0.00,20.00){\line(1,1){30.00}}
\put(20.00,0.00){\line(1,1){30.00}}
\put(4.00,27.00){\makebox(0,0)[rb]{$A$}}
\put(23.00,46.00){\makebox(0,0)[rb]{$C$}}
\put(46.00,23.00){\makebox(0,0)[lt]{$D$}}
\put(26.00,3.00){\makebox(0,0)[lt]{$E$}}
\put(27.00,27.00){\makebox(0,0)[lb]{$B$}}
\put(5.00,25.00){\circle*{1.00}}
\put(25.00,5.00){\circle*{1.00}}
\put(25.00,25.00){\circle*{1.00}}
\put(25.00,45.00){\circle*{1.00}}
\put(45.00,25.00){\circle*{1.00}}
\end{picture}
\end{center}
\caption{}
\label{figV-3-2}
\end{figure}
then this eigenvalue is $\varphi_D/\varphi_A$. It is clear that
$$
{\varphi_D\over \varphi_B}={\varphi_{A'}\over \varphi_{B'}}
$$
---the ratios of values $\varphi_{\ldots}$ in the triangle $DBE$ are
the same as in $A'B'C'$. Thus,
\be
{\varphi_D\over \varphi_A}={\varphi_{A'}\over \varphi_{B'}}
{\varphi_B\over \varphi_A}={\varphi_{A'}\over \varphi_A} \cdot
{-a(\varphi_A/\varphi_{A'})+1 \over (\varphi_A/\varphi_{A'})-a}
\label{V-3-6}
\ee
(we have used (\ref{V-3-3}). A similar relation can be written out
for the translation through one lattice period in {\em upward\/}
direction in the sense of Figures \ref{figV-3-1} and~\ref{figV-3-2},
namely
\be
{\varphi_C\over \varphi_E}=
{\varphi_{B'}\over \varphi_B} \cdot
{-\tilde a(\varphi_B/\varphi_{B'})+1 \over
(\varphi_B/\varphi_{B'})-\tilde a},
\label{V-3-7}
\ee
where one has to substitute the expression (\ref{V-3-3})
for $\varphi_B/\varphi_{B'}$.

It is clear that the ``dispersion relations'' of type
(\ref{V-3-6}--\ref{V-3-7}) survive also for a string ``created
by a particle'', if we substitute the eigenvalue of transfer matrix~$T$
instead of $\varphi_{A'}/ \varphi_A$, and the eigenvalues
of translation operators instead of $\varphi_D/ \varphi_A$
and $\varphi_C/ \varphi_E$. As for the multi-string states,
all of the eigenvalues are obtained for them as products
of corresponding eigenvalues for each string.

\section{Discussion}
\label{sec-V-discussion}

We have shown in this paper that the string---particle ``marriage''
 from paper~\cite{IV} makes possible a simple and clear construction
of at least some multi-string states. Recall that, from all
the corresponding multi-particle states, we only could
explicitely construct some two-particle states~\cite{I},
with an additional restriction that could be formulated
as ``the total momentum of two particles is zero''.
As for the present paper, the momenta of ``particles'' generating
the multi-string states of Sections \ref{secV-1} and~\ref{secV-2}
can change independently.

These states have been constructed for the infinite kagome lattice.
We have to recognize that constructing such states on a finite
lattice remains an open problem.

It is also unclear how to combine the results
of Sections \ref{secV-1} and~\ref{secV-2}, i.e.\ construct
such states with string ``creation'' and ``annihilation'' where
the total number of ``creating'' and ``annihilating'' particles
would be more than two. Note that in Section~\ref{secV-1} we use
the symmetrized product of one-particle amplitudes, while
in Section~\ref{secV-2}---the antisymmetrized one.

Concerning the dispersion relations of Section~\ref{V-sec-disp},
let us remark that perhaps there are too many of them.
It is probably caused by the fact that, for now, we managed
to construct not all one-particle and/\allowbreak or one-string states.


\begin{thebibliography}{99}

\bibitem{I}
I.G.~Korepanov, {\it Some eigenstates for a model associated with
solutions of tetrahedron equation}, solv-int/9701016, 7p.

\bibitem{II}
I.G.~Korepanov, {\it Some eigenstates for a model associated with
solutions of tetrahedron equation. II.~A bit of algebraization},
solv-int/9702004, 8p.

\bibitem{III}
I.G.~Korepanov, {\it Some eigenstates for a model associated with
solutions of tetrahedron equation.
III.~Tetrahedral Zamolodchikov algebras and perturbed strings},
solv-int/9703010, 7p.

\bibitem{IV}
I.G.~Korepanov, {\it Some eigenstates for a model associated with
solutions of tetrahedron equation.
IV.~String---particle marriage},
solv-int/9704013, 6p.

\end{thebibliography}
\end{document}